\documentclass[aps,pre,superscriptaddress,showpacs,twocolumn]{revtex4}
\usepackage{graphicx}
\begin{document}
\title{Strongly nonequilibrium flux flow in the presence of perforating submicron holes}
\author{D. Babi\'{c}}
%\altaffiliation[Corresponding author ] {}
%\email{dbabic@phy.hr}
\affiliation{Department of Physics, Faculty of Science,
University of Zagreb, Bijeni\v{c}ka 32, HR-10000 Zagreb, Croatia}
\author{J. Bentner}
\affiliation{Institut f\"{u}r experimentelle und angewandte Physik,
Universit\"{a}t  Regensburg, D-93025 Regensburg, Germany}
\author{C. S\"{u}rgers}
\affiliation{Physikalisches Institut and DFG Center for Functional Nanostructures
(CFN), Universit\"at Karlsruhe, D-76128 Karlsruhe, Germany}
\author{C. Strunk}
\affiliation{Institut f\"{u}r experimentelle und angewandte Physik,
Universit\"{a}t  Regensburg, D-93025 Regensburg, Germany}
%
%\date{\today}
%
%
%
\begin{abstract}
We report on the effects of perforating submicron holes on the vortex dynamics of amorphous Nb$_{0.7}$Ge$_{0.3}$ microbridges in the strongly nonequilibrium mixed state, when vortex properties change substantially. 
In contrast to the weak nonequilibrium - when the presence of holes may result in either an increase 
(close to $T_c$) or a decrease (well below $T_c$) of the dissipation, in the strong nonequilibrium 
an enhanced dissipation is observed irrespectively of the bath temperature. Close to $T_c$ this enhancement 
is similar to that in the weak nonequilibrium, but corresponds to vortices shrunk due to the Larkin-Ovchinnikov mechanism. At low temperatures the enhancement is a consequence of a weakening of the flux pinning by the holes in a regime where electron heating dominates the superconducting properties.

\end{abstract}
\pacs{74.78.Na, 74.40.+k, 74.25.Qt}
%
%74.78.Na Mesoscopic and nanoscale systems
%74.78.Db Low-Tc films
%74.25.Qt Vortex lattices, flux pinning, flux creep
%
\maketitle

\section{Introduction}

After more than a decade of the domination of high-$T_c$ compounds in superconductivity research, the interest for conventional type II superconductors has recently been renewed.
This trend is related to the necessity of a better understanding of vortex-motion phenomena rather than to improving the pinning in materials for high-current applications, since the complexity of the mixed state of high-$T_c$
superconductors has prevented a deeper insight into basic principles of vortex transport. 
A number of classical, low-$T_c$ alloys can nowadays be fabricated in the form of 
homogeneous thin films of highly reproducible superconducting properties and, furthermore,
can be shaped using standard methods for sample structuring - such as electron-beam lithography.
Some of these systems, for example amorphous Nb$_{1-x}$Ge$_x$, have a very weak pinning over a large portion of the magnetic field ($B$) vs temperature ($T$) plane \cite{kes1}, being hence suitable for studies on fundamental, pinning-independent properties of vortex dynamics. In turn, with such samples one can combine a simple vortex matter (characteristic of classical superconductors), an extended $(B,T)$ range of weak or negligible pinning (a property of high-$T_c$ materials) and artificially-introduced elements for influencing vortex motion locally.

Our recent investigation of vortex transport in amorphous Nb$_{0.7}$Ge$_{0.3}$ microbridges \cite{vnoise,ffi,eqholes}
has utilised the above-outlined advantageous properties of this material. We have obtained a quantitative understanding
of several mechanisms related to vortex dynamics in general, which has been used to
deal with some aspects of the role of small (of the order of the magnetic-field penetration depth $\lambda$)  perforating holes on the vortex-motion dissipation. 
In this work we concentrate on the dissipation in the presence of holes when the induced electric field
is thus strong that vortex cores suffer substantial changes. 
More precisely, we refer to the strongly nonequilibrium regime of vortex motion, the appearance of which is preceded
by nonlinear voltage-current characteristics of an origin different from depinning phenomena.
The physics underlying these nonlinearities is either vortex-core shrinking close to $T_c$, proposed theoretically by 
Larkin and Ovchinnikov (LO) \cite{lo}, or vortex-core expansion due to electron heating above 
the bath temperature \cite{kunchur},
depending on the experimental conditions. The above effect has been identified experimentally in a number of superconductors ranging from simple metals (Al, Sn, In) \cite{mus,klein} to high-$T_c$ materials \cite{doet},
including amorphous Nb$_{1-x}$Ge$_x$ \cite{vnoise,ffi,note1}. 
To our knowledge, the modification of the dissipation in this regime by perforating holes  has so far not been investigated, presumably because both the nature of the strongly nonequilibrium state and the effects of holes on vortex transport in the weak nonequilibrium were until recently not well understood. Having these topics analysed in 
Refs.\cite{ffi} and \cite{eqholes}, respectively, we now have a basis for turning to the present subject.  

% Fig.1
%
\begin{figure}
\includegraphics[width=75mm]{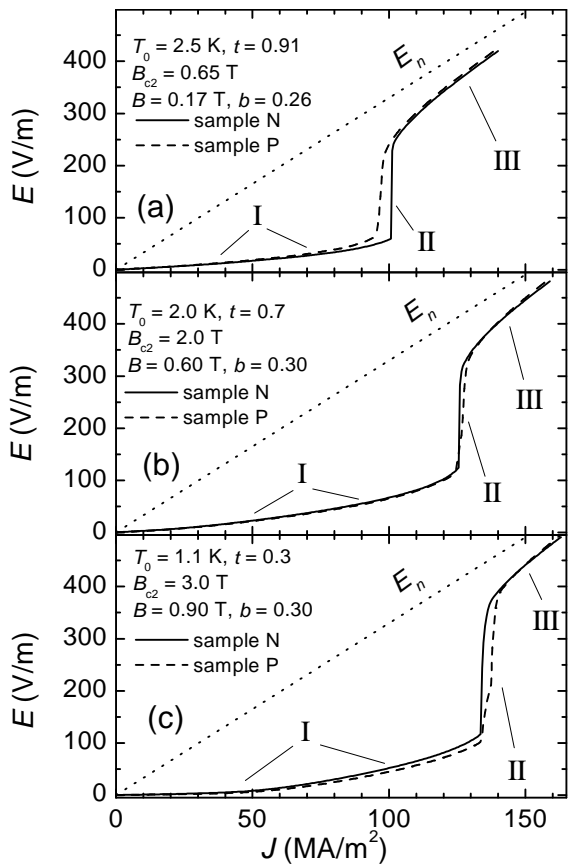}
\caption{$E(J)$ curves of sample N (solid lines) and sample P (dashed lines) for three bath temperatures
$T_0$ and the corresponding $B_{c2}(T_0)$, as indicated, and for $b = 0.26 -0.30$. Regimes I and III correspond to  the weak and strong nonequilbrium, respectively, with regime II in between. The dotted line represents the normal-state
electric field $E_n = \rho_n J$.}
\end{figure}

The main features of the overall dissipation can be seen in the electric field vs current density characteristics $E(J)$ in Fig.1, discussed in more detail later, where the results for three characteristic reduced temperatures  $t=T_0/T_c$ are shown ($t=0.91, 0.70, 0.40$). Here $T_0$ denotes the bath temperature, which is assumed to coincide with the phonon temperature. The reduced magnetic field $b = B/B_{c2}$ is $0.26 - 0.30$ ($B_{c2}$ is the equilibrium upper critical magnetic field), and the solid and dashed lines represent the $E(J)$ of a nonperforated (sample N) and a perforated 
(sample P) microbridge, respectively. There are three dynamic regimes, denoted by I, II and III, related to the specific properties of moving vortex cores. Regime I corresponds to the {\it weak nonequilibrium} (WN),
where vortex cores maintain the static properties, i.e. their radius is set by the equilibrium coherence length $\xi (T_0)$. As shown in Ref.\cite{eqholes}, in this case the presence of holes results in: (a) an increase of the dissipation close to $T_c$ due to a local modulation of the current density around the holes; (b), (c) a reduction of the dissipation arising from flux pinning by the holes, which is stronger as $T_0$ is lower. In regime III, $E$
is not much smaller than the normal-state electric field $E_n = \rho_n J$, where $\rho_n$ is the normal-state resistivity, which signifies dramatic changes in vortex cores. This is the 
regime of the {\it strong nonequilibrium} (SN),
the physics of which is discussed in Ref.\cite{ffi}. We shall return to this regime later, but at this point we note that, at all three characteristic temperatures, for sufficiently high $J$ the holes invariably {\it enhance} the dissipation - in contrast to regime I, and this is the central topic of this article.
The intermediate regime II is at low $b$ manifested by the presented steep jump - called flux-flow instability (FFI), whereas at higher $b$ this transition is smooth.

\section{Groundwork for the analysis}

In this Section we briefly summarise the results of Refs.\cite{ffi} and \cite{eqholes}, which are required for
addressing the SN vortex transport in the presence of perforating holes. 

\subsection{Experiment}

%Fig.2
%
\begin{figure}
\includegraphics[width=65mm]{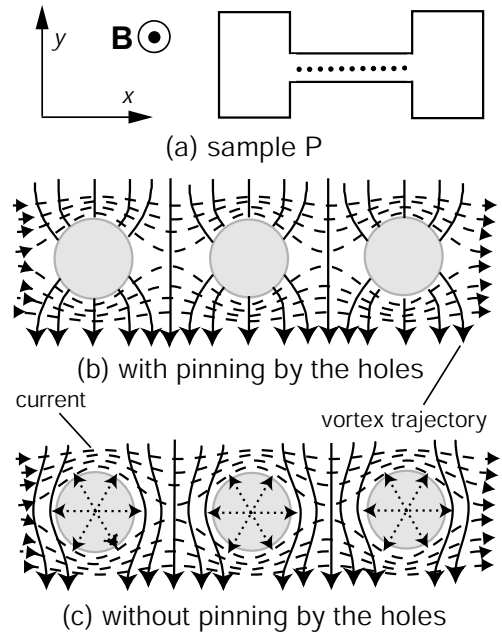}
\caption{(a) A sketch of sample P (not to scale) and the designation of the directions.
(b) The modulation of $J$ (arrowed dashed lines) around the holes and the vortex trajectories 
(arrowed solid lines) without
pinning by the holes. (c) The change of the vortex trajectories when the holes pin vortices, thus
exerting a repulsive force (radially-pointing arrowed dotted lines) upon the other vortices.}
\end{figure}

The vortex-motion dissipation has been studied by measurements of the {\it dc} $E(J)$ of two microbridges (samples N and P) between large contact pads, deposited simultaneously onto an oxidised Si substrate through the masks prepared by electron-beam lithography. Both microbridges are $L=210$ $\mu$m long, $W=5$ $\mu$m wide and $d=20$ nm thick. While sample N is a plain microbridge, sample P is perforated with a line of equidistant holes 800 nm in diameter, with the centre-to-centre separation of 1.2 $\mu$m, as sketched in Fig.2(a). An applied magnetic field ${\bf B}$ is oriented
perpendicularly to the film plane, a transport current is passed along the bridge (in the $x$ direction),
vortices traverse the sample in the $y$ direction, thus the measured $E$ is in the $x$ direction.
Both samples have the same $T_c = 2.75$ K, the Ginzburg-Landau parameters are $\xi(0)=6.8$ nm and $\lambda(0) = 1.15$ $\mu$m, and $\rho_n = 3.3$ $\mu \Omega$m 
(in the temperature range of interest changes at the level of 0.01 \%). 

Clearly, in sample P the current density is spatially modulated around the holes, hence if we want to compare its $E(J)$ with that for sample N we have to find the ratio of the {\it average} current density $J_P$ in sample P and the uniform current density $J_N$ in sample N. This has been done by comparing the resistances above $T_c$, which match perfectly if $J_P = 1.23 J_N$. Using this relationship we can calculate and compare $E_N (J_N)$ and $E_P(J_P)$ for samples N and P, respectively, and use this notation when referring to to the $E(J)$ of the two samples specifically. $E_N(J_N)$ and $E_P(J_P)$ were measured at constant $B$ and $T$ by sweeping the applied current up at a rate of 10 nA/s, which was slow enough to avoid any detectable dependence of $E$ on the sweep rate. 
 
\subsection{Effect of holes on vortex transport in weak nonequilibrium}
\label{holeq}

In the WN regime, vortex cores preserve their static properties and vortex transport corresponds to a Lorentz-force-induced motion of normal cylinders of a radius $\sim \xi (T_0)$, each carrying a quantum $\phi_0$ of the magnetic flux. This force is opposed by a viscous damping $\eta {\bf u}$, where $\eta$ is the viscosity coefficient and ${\bf u}$ the vortex velocity, and by the pinning force ${\bf F}_p$ (if present). The equation of motion for vortices created by a magnetic field ${\bf B} = B {\bf \hat{z}}$, $| {\bf \hat{z}} | = 1$, is (per unit vortex length)
$m \dot{{\bf u}} = {\bf F}_d - \eta {\bf u} - {\bf F}_p$,
where ${\bf F}_d = \phi_0 {\bf J} \times {\bf \hat{z}}$ is the driving force, $m$ the effective vortex mass \cite{mass},
and the $i$th component of the vortex acceleration is given by $\dot{u}_i = {\bf u} \cdot \nabla u_i$. If ${\bf J}$ is uniform and if there are no other forces on vortices, the left-hand side of the above equation equals zero and we obtain the well-known equation of vortex motion in homogeneous superconductors. The electric field induced by vortex motion is
${\bf E}({\bf r}) = {\bf u}({\bf r}) \times {\bf B}({\bf r}) = 
\phi_0 n ({\bf r}) [ {\bf u}({\bf r}) \times {\bf \hat{z}}]$,
where $n = B/ \phi_0$ is the vortex density. Again, for uniform ${\bf u}$ the coordinate dependences in the above expressions can be omitted. The simplest case is ${\bf F}_p = 0$, which is in our samples satisfied close to $T_c$ (typically, for $t > 0.9$) over most of the range $B < B_{c2}$, when the dissipation for uniform ${\bf u}$
(sample N) agrees very well with the LO theory \cite{lo} of ohmic flux-flow resistivity $\rho_f$.

In the presence of holes the current density is spatially modulated in their vicinity
(Figs.2(b) and 2(c), arrowed dashed lines), which leads to a nonuniform
${\bf F}_d$ and the relevance of the term $m \dot{{\bf u}}$. As discussed in detail in Ref.\cite{eqholes}, with ${\bf F}_p = 0$ and no pinning by the holes the modulation of ${\bf J}$ causes a chanelling of the vortex trajectories
through the holes (Fig.2(b), arrowed solid lines) and an enhancement of the measured dissipation
(Fig.1(a), regime I). 

With decreasing $T_0$ the pinning strength of the holes increases,
which introduces an additional term on the right-hand side of the equation of motion, i.e. a repulsion
of the vortices pinned by the holes and the other vortices [as sketched by the arrowed
dotted lines in Fig.2(c)]. This changes the vortex trajectories as indicated, and results in
a reduction of the dissipation [regime I in Figs.1(b) and 1(c)], an effect that is stronger as $T_0$ is lower.

It is worthwhile to note that the results outlined above address the influence of a single hole on vortex transport more directly than those referring to two-dimensional arrays of holes in samples with stronger intrinsic pinning than in Nb$_{0.7}$Ge$_{0.3}$. In these experiments the main effects arise from an interplay of the material pinning in the interstitial area and the matching effects between the vortex lattice and the regular array, see e.g. Ref.\cite{mosch}, which can even disappear below $t \sim 0.9$. 

\subsection{The nature of strongly nonequilibrium flux flow} 
\label{nature}

The nature of the regime denoted by III in Fig.1 was until recently not completely understood. LO predicted a jump in $E(J)$, i.e. FFI, as a consequence of a decrease of $\eta$ caused by a shrinking of vortex cores according to 
$\xi(T,E) = \xi (T,0) / [ 1 + (E/E_i)^2]$, where $E_i = u_i B$ is a characteristic electric field corresponding to a vortex velocity $u_i$. The existence of the FFI has been verified in a number of experiments, but most of these have relied on simply identifying $E_i$ from the position of the jump, without addressing the regime
$E_i < E < E_n$ in detail. Furthermore, the LO mechanism turned out to be viable only close to $T_c$, failing to account for the SN state at low $T_0$. 
 
A quantitative treatment of $E(J)$ in Ref.\cite{ffi} showed that close to $T_c$ the LO vortex-core shrinking
was a valid mechanism up to at least $b \sim 0.75$, which was considerably higher than the theoretical prediction of 
Bezuglyj and Shklovskij \cite{bs}, who estimated that heating effects should be significant above $b \sim 0.4$. For the present discussion it is important that the LO description holds for $E_N(J_N)$ in regime III at $t=0.91$  over the entire range of measurement and analysis ($0.15 \lesssim b \lesssim 0.75$). 

Far below $T_c$ neither the FFI nor the SN $E(J)$ can be described by the LO approach. The low-$T_0$ FFI was addressed by Kunchur \cite{kunchur}, who assumed that the nonlinearities 
in $E(J)$ originated in heating of electrons to a temperature $T^* > T_0$. 
This hypothesis sets $T^*$ rather than $T_0$ as the relevant temperature for superconducting parameters, which
results in an expansion of vortex cores, according to $\xi (T^*) > \xi(T_0)$, and a consequent decrease of the $B_{c2}$.
The same idea was used in Ref.\cite{ffi} to quantitatively describe the SN $E(J)$ (regime III) by a replacement $B_{c2}(T^*) \rightarrow B_{c2}(T_0)$. This analysis also led to identifying a nonequilibrium phase boundary $E_c(B,T_0)$ as an electric field at which the heating destroyed the superconductivity. 

Hence, the generic properties of the SN flux flow in our samples are understood sufficiently well to discuss
the modifications introduced by the holes.

\section{Analysis and discussion}
 
\subsection{Temperatures close to $T_c$}
\label{close}

We first consider the results for $T_0$ close to $T_c$, a representative of which is shown in Fig.1(a). 
This and the other results of this Section refer to $T_0 = 2.5$ K, i.e. $t=0.91$, where 
$E_N(J_N)$ (measured for $0.15 \lesssim b \lesssim 0.75$, with $B_{c2} \approx 0.65$ T) agrees remarkably well with the LO model of flux flow \cite{lo} for all three regimes I, II and III. More precisely, a single theoretical curve, which in the limit $E \ll E_i$ reduces to ohmic $\rho_f = E/J$, describes the entire $E_N(J_N)$.  
Noteworthy, for these data we have found no signature of intrinsic pinning, 
i.e. in the limit $J \rightarrow 0$ the results agree well with the theoretical $\rho_f$ calculated for
depinned vortices. The reader may consult 
Ref.\cite{ffi} for details of the comparison of the experiment and theory, whereas here we recall 
the main results only. 

The nonlinearities of $E_N(J_N)$ are in a one-to-one correspondence to the shrinking
of vortex cores mentioned in Section \ref{nature}. The characteristic electric field $E_i$ has been found
to be linear with $B$, as predicted by LO, from which we have extracted the characteristic vortex velocity $u_i \approx 305$ m/s. When $u$ exceeds $u_i$ considerably, the shrinking stops at $\tilde{\xi} = \xi (t) (1 - t)^{1/4}$, which defines SN vortex radius. In our case ($t=0.91$), $\tilde{\xi} \approx 0.55 \xi$ and
the SN flux flow refers to vortices of approximately three times smaller area than in equilibrium.
Therefore, in regimes I and III the vortex radii are essentially independent of $u$, being $\xi$ and 
$\tilde{\xi}$, respectively, while in regime II they depend on $u$ strongly \cite{note2}. 

The current-density modulation around the holes, sketched in Figs.2(b) and 2(c), in the WN (regime I)
causes an increase of the dissipation in comparison to that with no holes \cite{eqholes}. The vortices follow the trajectories depicted in Fig.2(b), which is possible because the flux pinning by the holes is weak due to 
the large $\lambda$ close to $T_c$ \cite{eqholes,mosch,take}. The vortex velocity obviously varies in space, i.e.
it is larger in the regions of larger $J$, and the net result is $E_P > E_N$. 

We see in Fig.1(a) that this behaviour continues in regimes II and III. Hence, close to $T_c$ the current modulation around the holes always enhances the dissipation, irrespectively of the particular dynamic radius of vortex cores. In region III the situation is simpler than in region II,  because in this case the vortices have already shrunk to their minimum size - which is therefore constant over the entire sample. On the other hand, in regime II the vortex-core size varies considerably with $u$ and consequently
depends on the vortex position.

% Fig.3
%
\begin{figure}
\includegraphics[width=75mm]{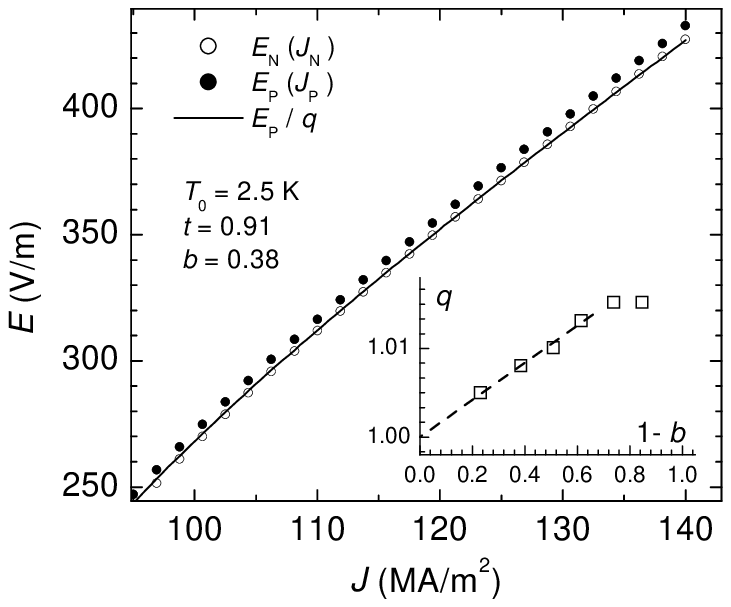}
\caption{$E_N(J_N)$ (open circles) and $E_P(J_P)$ (solid circles) in the strongly nonequilbrium state
(regime III) at $T_0 = 2.5$ K ($t=0.91$) and $b=0.38$. The solid line represents $E_P / q$ with $q=1.013$.
Inset: A plot of $q$ against $1-b$, showing that the values of $q$ extrapolate to unity in the limit $b \rightarrow 1$, 
as indicated by the dashed line.}
\end{figure}

The dissipation enhancement in regime III can be approached quantitatively. 
If the modulation of $J$ is the main cause of $E_P > E_N$ for vortices of a radius $\tilde{\xi}$, thein it should be
possible to obtain $E_P(J_P) = q E_N(J_N)$ with $q \geq 1$ being a constant over a range of $J$.
This is the case indeed, as demonstrated in Fig.3 for $b=0.38$ where the agreement for
$J \gtrsim 110$ MA/m$^2$ is remarkable. 
$E_N(J_N)$ and $E_P(J_P)$ are represented by the open and solid circles, respectively, while the solid line is
$E_P /q$ plotted using $q=1.013$.
Although the differences are small - around 1 \%, they are of the order of 1 mV and are thus reliably observable. This procedure can be carried out for all the $E(J)$ to extract the $b$ dependence of $q$. The result is shown 
in the inset to Fig.3, where $q$ is plotted against $1-b$. Obviously, the values of $q$ extrapolate to unity for $b \rightarrow 1$,
thus satisfying the condition that the two dissipations must be the same in the normal state.   

A quantitative treatment of  regime II would require a detailed numerical modelling, which is outside the scope of this paper, thus we restrict ourselves to qualitative considerations. Besides the observed $E_P > E_N$, 
another feature is discernible: regime II is for sample N always narrower (and the FFI, if present, sharper), which
complies with the picture of locally-dependent vortex velocity in sample P. Namely, in this case the measured $E$ corresponds to an integration over a distribution of $\xi(u)$, resulting in a broader range of the strong nonlinearity, here denoted as regime II.  

In conclusion to Section \ref{close}, our results and analysis suggest that the current-density modulation around
perforating holes of a diameter $\sim \lambda(0)$ may increase the dissipation irrespectively of the dynamic 
properties of vortex cores. This can occur provided the hole pinning is sufficiently weak, i.e. in conditions of a large $\lambda (T_0 \rightarrow T_c)$. The SN dissipation enhancement by the holes can be described by a 
$b$-dependent factor $q = E_N/E_P$. By implication, the system remains in thermal equilibrium at $T_0$, and the increase of $E$ around the holes is solely a consequence of the modulation of $J$.  

\subsection{Temperatures well below $T_c$}
\label{lowt}

As already mentioned in Section \ref{holeq}, perforating holes pin equilibrium vortices most efficiently if $\lambda$ is of the order or of the hole diameter, as calculated theoretically \cite{take} and demonstrated experimentally \cite{eqholes,mosch}. Therefore, the weak pinning by the holes close to $T_c$ is a consequence of the large $\lambda (T_0 \rightarrow T_c)$, and, as shown in Section \ref{close}, the same holds for vortices which undergo the LO core shrinking. 

In the WN at low $T_0$ the holes are efficient pinning centres, which can be understood in terms of a smaller $\lambda(T_0)$, leading to a suppression of the dissipation \cite{eqholes}.
This behaviour is seen in Figs.1(b) and 1(c) (regime I), where the dissipation suppression is stronger as $T_0$ is lower and $\lambda$ consequently smaller. As argued in Ref.\cite{eqholes}, the reason for the dissipation reduction is not only in some vortices being immobilised by the holes but also in changing the vortex trajectories as depicted in Fig.2(c), so that the mobile vortices pass through the regions of smaller current density, between the holes. The vortex trajectories change because the vortices pinned by the holes repel the mobile vortices, as sketched by the arrowed dotted lines in Fig.2(c). The net reduction of the dissipation - {\it originating from the both effects mentioned above}, decreases monotonically with $b$. More precisely, this behaviour is described quantitatively by a ratio $p=P_P / P_N$ of the power density  $P_\gamma = \int E_\gamma dJ_\gamma$ dissipated in samples P and N (as denoted by the subscripts $\gamma = P,N$) in regime I. For the $E(J)$ measured at $T_0 = 1.1$ K, $p$ increases from $\sim 0.77$ at $b=0.2$ to $\sim 0.98$ at $b=0.83$. The same effect appears for $T_0 = 2.0$ K as well, but in this case $p \rightarrow 1$ already around $b \sim 0.6$. 

%Fig.4
%
\begin{figure}
\includegraphics[width=75mm]{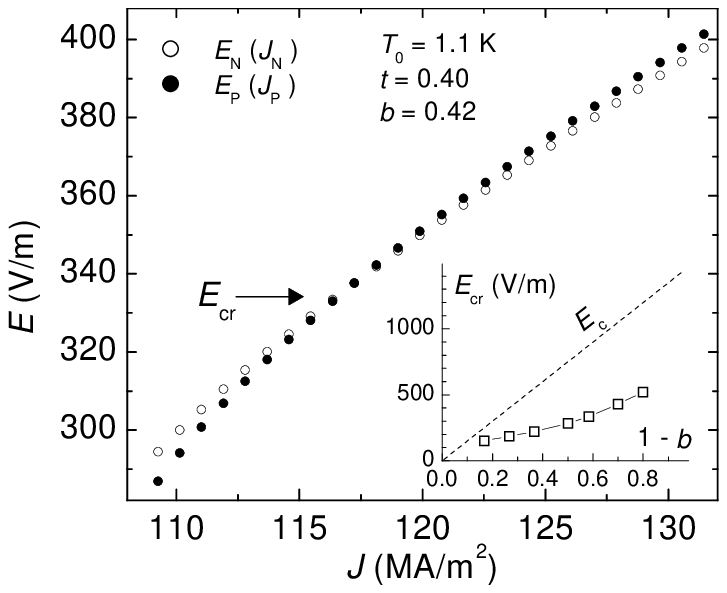}
\caption{$E_N(J_N)$ (open circles) and $E_P(J_P)$ (solid circles) in the strongly nonequilibrium state (regime III)
at $T_0 = 1.1$ K ($t=0.40$) and $b=0.42$. At $E_{cr} \approx 335$ V/m  there is a crossover from $E_N > E_P$,
characteristic of regimes I and II, to $E_N < E_P$. Inset: The crossover field $E_{cr}$ against $1-b$ (open squares), and $E_c \propto 1-b$, represented by the dashed line, that marks a transition to the normal state by the overheating (as inferred in Ref.\cite{ffi} for sample N).}
\end{figure}

If we now compare $E_N$ and $E_P$ in regime II, we see in Fig.1 that, in contrast to the situation close to $T_c$,
$E_P < E_N$ for $T_0 = 2.0$ K and 1.1 K, again with the difference being larger at lower $T_0$. This implies that the holes continue to act as efficient pinning sites even when the nonequilibrium effects start to take place. However, when the SN  fully develops, i.e. in regime III, we observe a crossover to $E_P > E_N$ at an electric field $E_{cr} (b)$, as demonstrated in Fig.4 in an expanded scale for $T_0 = 1.1$ K and $b=0.42$. 

For $T_0 = 1.1$ K the crossover appears over a wide $b$ range, thus we can study $E_{cr}$ as a function of $b$.
The determined values of $E_{cr}$ are in the inset to Fig.4 plotted against $1-b$ by the open squares.  
The dashed line represents $E_c = E_{c0} (1-b)$ with 
$E_{c0} \approx 1500$ V/m, deduced in Ref.\cite{ffi} for sample N and interpreted as a dynamic phase boundary between the SN mixed state and the normal state. Therefore, in the SN the holes enhance the dissipation even if they pin equilibrium vortices efficiently, and this experimental fact is the subject of the discussion below.      

At low $T_0$, the SN corresponds to electron heating to a temperature $T^* > T_0$
(see Section \ref{nature}), leading to a decrease of $B_{c2}$ and eventually to a transition to the normal state when $b^* = B/B_{c2}(T^*) = 1$ (or, equivalently, $E=E_c$). The heating also causes an increase of $\lambda$, which implies a weakening of the pinning by the holes in comparison to the equilibrium and WN conditions. In this process the vortex trajectories depicted in Fig.2(c) are expected to change to those in Fig.2(b). In other words, if no vortices are pinned by the holes the spatially modulated $J$ governs the vortex motion and we could again expect an enhancement of the dissipation - which is observed indeed. 

% Fig.5
%
\begin{figure}
\includegraphics[width=75mm]{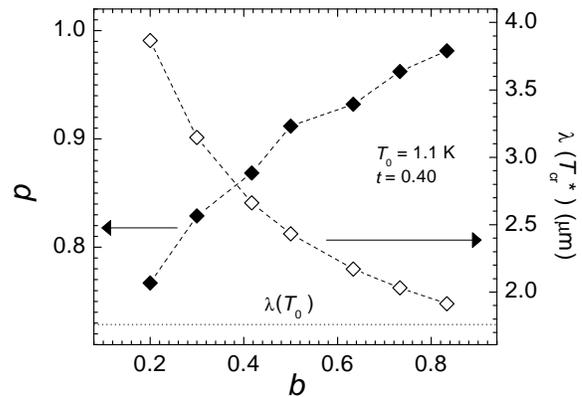}
\caption{The parameter $p$, describing the pinning strength of the holes in regime I (solid diamonds, left-hand scale), and $\lambda (T^*_{cr})$ (open diamonds, right-hand scale) as a function of $b$, at $T_0 = 1.1$. The former quantity has been taken from 
Ref.\cite{eqholes}, whereas the latter has been calculated in the approximation of a temperature-independent $\kappa$
in the spirit of the electron-heating model of Ref.\cite{ffi}. The equilibrium $\lambda(T_0)$ is indicated by the dotted line.}
\end{figure}

We can inspect the crossover at $E_{cr}$ more closely. From the heating characteristics $T^*(E,B)$ inferred in 
Ref.\cite{ffi} for sample N it is possible to calculate $T^*_{cr} = T^* (E_{cr})$ and $b^*_{cr} = b^* (T_{cr})$.
We find that $T^*_{cr}/T_c(B)$ and  $b^*_{cr}$ are both $\sim 0.96 - 0.98$, where 
$T_c(B)$ corresponds to the equilibrium $B_{c2}(T)$. Although $b^*_{cr}$ and $T^*_{cr}$ are thus close to unity,
the range $E_{cr} < E < E_c$ is not necessarily small. This is a consequence of the form 
\cite{lo} of $J = \sigma_n [ 1 + \alpha (1-b)]E$ and $b^*(E)$ that describe
the $E(J)$ by a replacement $b \rightarrow b^*$ (with $\alpha \approx 3$) \cite{ffi}. 
In the approximation of a temperature-independent Ginzburg-Landau parameter $\kappa$, in Ref.\cite{ffi} found to be applicable, we can calculate $\lambda (T^*_{cr})$ by inserting $\lambda = 1.63 \kappa \xi$ into $B_{c2} = \phi_0 / 2 \pi \xi^2$. The result of this calculation is displayed in Fig.5, where $\lambda(T_{cr}^*)$ is plotted vs $b$ together with the parameter $p$ defined earlier and determined in Ref.\cite{eqholes}. 

The $b$ dependence of $p$ is mainly of a dynamic origin, since the vortices incoming to the holes push those pinned therein, this effect being stronger as the vortex density grows \cite{eqholes}. At low vortex density the majority of vortices either stay pinned by the holes for a prolonged time or are scattered to pass through the regions of low current density. In this case the heating must be rather strong for the crossover to occur, for instance $\lambda(T^*_{cr})$ at $b=0.2$ is close to
$\lambda(T_0 = 2.5 \; {\rm K}) \approx 3.8$ $\mu$m. 

The above explanation of the crossover in terms of an increase of $\lambda(T^*)$ in principle assumes $T^*$ constant over the entire sample, i.e. a complete thermal equilibration within the electron system. However, $T^*(E)$ can be spatially dependent through the coordinate dependence of the local $E$. Thus, we cannot rule out a possibility of a creation of hot spots at the apices of the holes, where $E$ is the largest if there is no hole pinning [see Fig.2(b)]. Such hot spots may be a cause of the observed widened low-$T_0$ FFI, such as that in Fig.1(c), but we believe that they
do not affect the above picture substantially.

\section{Conclusions}
  
We present and discuss  measurements of the voltage-current characteristics in the mixed state of two amorphous Nb$_{0.7}$Ge$_{0.3}$ microbridges, with and without a line of perforating submicron holes, with emphasis on the strongly nonequilibrium flux flow. In the strongly nonequilibrium regime of vortex motion we find an enhancement of the dissipation by the holes both close to $T_c$ and well below $T_c$. The 
strong equilibrium close to $T_c$ is related to vortices shrunk due to the Larkin-Ovchinnikov mechanism, the dissipation enhancement is similar to that observed in the weak nonequilibrium and originates in a spatially nonuniform current distribution in the vicinity of the holes. The effect of the current-density modulation is dominant because of a negligible flux pinning by the holes, which is a consequence of the magnetic-field penetration depth $\lambda$ being much larger than the hole diameter. At low temperatures $\lambda$ decreases and the holes pin equilibrium vortices, which reduces the net dissipation. This effect, however, does not continue in the strong equilibrium, where a crossover to the opposite behaviour occurs. Since the strong nonequilibrium at low temperatures
most probably corresponds to electron heating to above the bath temperature, it is likely that the crossover is a consequence of an increase of $\lambda$ by the heating, accompanied by a reduction of the pinning strength of the holes. 

In the weak nonequilibrium, holes in the geometrical arrangement of our samples can either enhance or reduce the vortex-motion dissipation. In contrast, when the strong nonequilibrium fully develops the presence of holes always results in an increased dissipation. Together with Refs.\cite{ffi} and \cite{eqholes}, the above findings provide a quantitative understanding of the effects of
perforating holes on vortex transport over a wide range of possible dynamic forms of the mixed state.

This work was partly funded by the Deutsche Forschungsgemeinschaft within the Graduiertenkolleg 638, the Croatian Ministry of Science, Education and Sport (Project No. 119262) and the Bavarian Ministry of Science, Research and Art.    


\begin{thebibliography}{15}
\bibitem{note2} Since $E=uB$, the $E$ and $u$ dependences are equivalent.
\end{thebibliography}
\end{document}